\documentclass[fleqn,usenatbib]{mnras}

\usepackage{lmodern}
\let\la=\lesssim % for less than similar from newtxmath, not \la from mnras.cls

\usepackage[T1]{fontenc}
\usepackage{ae,aecompl}
\usepackage{graphicx}
\usepackage{subcaption}
\usepackage{float}
\DeclareRobustCommand{\VAN}[3]{#2}
\let\VANthebibliography\thebibliography
\def\thebibliography{\DeclareRobustCommand{\VAN}[3]{##3}\VANthebibliography}

\usepackage{graphicx}	% Including figure files
\usepackage{amsmath}
\usepackage{sidecap}	% Extra maths symbols
\usepackage{bigfoot}
\usepackage{verbatim}
\usepackage{xcolor}
\usepackage{multirow}
\usepackage{multicol}
\usepackage{booktabs}
\usepackage{url}

\def\850{850~$\mu$m}
\def\550{550~$\mu$m}
\def\350{350~$\mu$m}

\def\lg353{L$_{\rm 353GHz}$}
\def\ln857{L$_{\rm 857GHz}$}

\newcommand{\z}{z_s}

\usepackage{newtxtext,newtxmath}

\title[AGN fraction in high-z protocluster candidates]{The AGN fraction in high-redshift protocluster candidates selected by Planck and Herschel}

\author[Gatica, C. et al]{Caleb Gatica $^{1}$\thanks{E-mail: calgaticai@gmail.com},
Ricardo Demarco $^{1}$,
Herv\'e Dole$^{2}$, Maria Polletta$^{3}$, Brenda Frye$^{4}$, Clement Martinache$^{1}$,
\newauthor Alessandro Rettura$^{5}$
\\
$^{1}$ Departamento de Astronom\'ia, Facultad de Ciencias F\'isicas y Matem\'aticas, Universidad de Concepci\'on, Concepci\'on, Chile\\
$^{2}$ Universit\'e Paris-Saclay, CNRS, Institut d'Astrophysique Spatiale, 91405, Orsay, France \\
$^{3}$ INAF - Istituto di Astrofisica Spaziale e Fisica cosmica (IASF) Milano, via A. Corti 12, 20133 Milan, Italy \\
$^{4}$ Department of Astronomy/Steward Observatory, University of Arizona, 933 N. Cherry Avenue, Tucson, AZ 85721, USA\\
$^{5}$National Aeronautics and Space Administration Jet Propulsion Laboratory, Pasadena, California, USA
}

\date{Accepted XXX. Received YYY; in original form ZZZ}

\pubyear{2023}

\begin{document}
\label{firstpage}
\pagerange{\pageref{firstpage}--\pageref{lastpage}}
\maketitle

\begin{abstract}
    A complete understanding of the mass assembly history of structures in the universe requires the study of the growth of galaxies and their supermassive black holes (SMBHs) as a function of their local environment over cosmic time. In this context, it is important to quantify the effects that the early stages of galaxy cluster development have on the growth of SMBHs. We used a sample of Herschel/SPIRE sources of $\sim$ 228 red and compact Planck-selected protocluster (PC) candidates to estimate the active galactic nuclei (AGN) fraction from a large sample of galaxies within these candidates. We estimate the AGN fraction by using the mid-infrared (mid-IR) photometry provided by the WISE/AllWISE data of $\sim650$ counterparts at high redshifts. We created an AllWISE mid-IR colour-colour selection using a clustering machine learning algorithm and two {\it WISE} colour cuts using the 3.4 $\mu m$ (W1), 4.6 $\mu m$ (W2) and 12 $\mu m$ (W3) passbands, to classify sources as AGN. We also compare the AGN fraction in PCs with that in the field to better understand the influence of the environment on galaxy development. We found an AGN fraction of $f_{AGN} = 0.113 \pm 0.03$ in PC candidates and an AGN fraction of $f_{AGN} = 0.095 \pm 0.013$ in the field. We also selected a subsample of `red' SPIRE subsample with a higher overdensity significance, obtaining $f_{AGN} = 0.186 \pm 0.044$, versus $f_{AGN} = 0.037 \pm 0.010$ of `non-red sources', consistent with higher AGN fractions for denser environments. We conclude that our results point towards a higher AGN fraction in PCs, similar to other studies.
\end{abstract}


%%%%%%%%%%%%%%%%%%%%%%%%%%%%%%%%%%%%%%%%%%%%%%%%%%
%%%%%%%%%%%%%%%%% BODY OF PAPER %%%%%%%%%%%%%%%%%%

\section{Introduction}
\label{sectintro}

Galaxies in the universe are not randomly distributed in space; instead, they can be isolated (i.e. a field galaxy) or in gravitationally-bound structures, such as groups or galaxy clusters (e.g. \citealt{oort83, waldrop83}).

Observations show that cluster galaxies exhibit different properties than those in the field. The distinction between these two in the local universe is seen in their morphologies (e.g. \citealt{Oemler1974, Dressler1980, Butcher1984, Goto2003}), colours (e.g. \citealt{Kennicutt1983,Hogg2003}), star formation (e.g. \citealt{Wolf2009, Salerno2022, Finn2023, Qu2023}), and other properties (e.g. \citealt{Boselli2006, Cappellari2013,  PerezMillan2023}); and it is also seen at higher redshifts (e.g. \citealt{Poggianti99, Gobat2008, Wolf2009, Demarco10, Rettura10, Papovich2012, Bassett2013, Pintos-Castro2019, Castignani2020}).

This raises important questions about the differences between galaxies as a function of their environment and how their evolutionary paths vary over cosmic time from the first density fluctuations to local current structures. To answer these questions, it is necessary to study protoclusters (PCs) of galaxies, the progenitor structures of today's massive clusters, during the epoch of their formation \citep{muldrew15, overzier16, Muldrew2018}. Consequently, there is a need to identify and characterise PCs at high redshift at $z \sim 2-3$ (`cosmic noon'), which corresponds to the time in cosmic history when the peak of the SFR density of the Universe occurs \citep{madauDickinson14, Forster2020}.
 
In order to be able to provide a complete picture of galaxy evolution as large-scale structures assemble and develop, we must understand the simultaneous growth of galaxies and their supermassive black holes (SMBHs). The active growth of a SMBH in a galaxy is typically signalled during its most vigorous phases of mass accretion (active galactic nucleus; AGN). In this regard, there is evidence of a peak at $z \sim 2-3$ for high-luminosity AGN  \citep{hasinger05, fanidakis12}, cosmic BH accretion (e.g. \citealt{Croom2009, Delvecchio2014}) and space-density of quasars (e.g. \citealt{Brown2006, Richards2006}).

AGN activity in different environments has also been explored. Results include lower average AGN fractions in clusters at redshift $z<0.5$ \citep{Mishra2020} when compared to the field, no dependence of the optical AGN activity on environment in blue galaxies \citep{Miraghaei2020}, higher AGN fractions for massive galaxies than lower mass galaxies \citep{Pimbblet2013}, similar AGN fractions in clusters and the field at $0.5 < z < 0.9 $ \citep{Klesman2012}, and an increase of AGN fractions with redshift \citep{Eastman2007}. Nevertheless, it is still unclear whether or not the local environment of galaxies plays a significant role in the growth of galaxies and their SMBHs. To clarify this, a statistical study of the environment hosting AGN activity is required, and in particular, it is important to determine the occurrence of AGN in PCs and at different and higher redshifts.

Studies of high redshift PCs concluded that they exhibit higher fractions of AGN and star-forming galaxies compared to the field, as opposed to overdensities at lower redshifts which have lower fractions than the field (e.g. \citealt{overzier16}, a review).
For instance, AGN fractions measured in PC range between 2 and 20 times higher than in the field \citep{Lehmer2009, Lehmer2013, Digby2010, Krishnan2017}. Also \cite{Polletta2021} found similar results for AGN fraction ($=13\% \pm 6\%$) in a PC at $z=2.16$. Recently, \cite{Macuga19} found a PC at $z=2.53$ with an AGN fraction $\sim$2 times lower than in the field, indicating a lack of clarity regarding the AGN activity in PCs.
Further, all of these studies showing a larger AGN fraction are based on X-ray selected AGN (see \citealt{Casey2014} for a review) . This type of selection is biassed against highly obscured sources \cite[and references therein]{Hickox2018, Hatcher2021}. Therefore, to provide a complete picture, other methods must be used to select AGN.  

Comparing all of these studies is difficult, since they all present different methods for selecting AGN or AGN contribution, different sensitivity limits, and definitions of non-AGN host galaxies (see \citealt{Padovani2017} for review). Whether differences in AGN fractions are due to redshift evolution, observational biases of PC selected in different halo masses, or evolutionary states, variations in the general and systematic properties of PC depending on how they are selected or just individual PC-to-PC variations remains a crucial open question.

Large samples of PC candidates have been built using large photometric surveys that have mapped significant areas of the sky, and some effort has been made to characterise these kind of environments (e.g., \citealt{Chiang13, Umehata15, Lee16, shimawaka18, miller19}). Performing a larger census of galaxies, especially AGN, within PCs is crucial to understanding the physical processes involved and determining whether the environment of a forming galaxy cluster at high redshift can trigger or drive the growth of SMBHs in its member galaxy population.

The main goal of this study is to measure the AGN fraction in a large sample of PC candidates. We use the sample of 228 Planck-selected PC candidates found in \citealt[]{PlanckCollaboration2015} (hereafter Planck XXVII), which itself is drawn from a more general sample of the Planck list of high-redshift source candidates (PHZ, \citealt{PlanckXXXIX}). This sample was followed up by Herschel/SPIRE, and it is biassed towards highly star-forming regions.
We combined the Planck XXVII catalogue with data from the Wide-field Infrared Source Explorer (WISE; \citealt{wise10}) AllWISE data release, which has mapped the whole sky. Using WISE sources allows us to use a mid-IR method that selects both obscured and unobscured AGN \citep{stern12}. For this, we built a classifier that includes both a clustering machine learning algorithm and W1-W2-W3 colour cuts. With the classification of our sources, we were able to estimate AGN fractions in both PC members and field galaxies. 

This paper is organised as follows: in Section \ref{sec:data-cat} we describe our Planck XXVII sample and its WISE counterparts, along with the control sample needed to construct our classifier; in Section \ref{sec:agn-class} we present how we classify AGN sources with our classifier together with estimates of the method uncertainty; in Section \ref{sec:results} we present the measured AGN fractions and our comparison to previous results in the literature; in Section \ref{sec:discussion} we discuss our results; and in Section \ref{sec:conclusions} we summarised our results and present our conclusions.

\section{Data \& catalogue} \label{sec:data-cat}

In this work, we use a catalogue of Planck colour-selected sources from the \citealt{PlanckCollaboration2015} (Planck XXVII), which corresponds to a catalogue of high-redshift protocluster candidates. This sample has follow-up observations with Herschel/SPIRE and the sources detected at $>$ $3\sigma$ in the Herschel/SPIRE 350$\mu$m band will be referred to as "SPIRE sources". This sample is what we consider our main sample, and it is described in Section \ref{sec:planck27}.
To have higher resolution photometry than Herschel/SPIRE, we derive the AGN fraction using their WISE counterparts. The description of this sample is in Section \ref{sec:wise-count}.
 
Also, to create a classification scheme that selects AGN, we compiled a control sample that includes catalogues of AGN (see Section \ref{sec:AGN}) and non-AGN sources (see Section \ref{sec:noAGN}).

\begin{figure}
    \centering
    \includegraphics[width=\columnwidth]{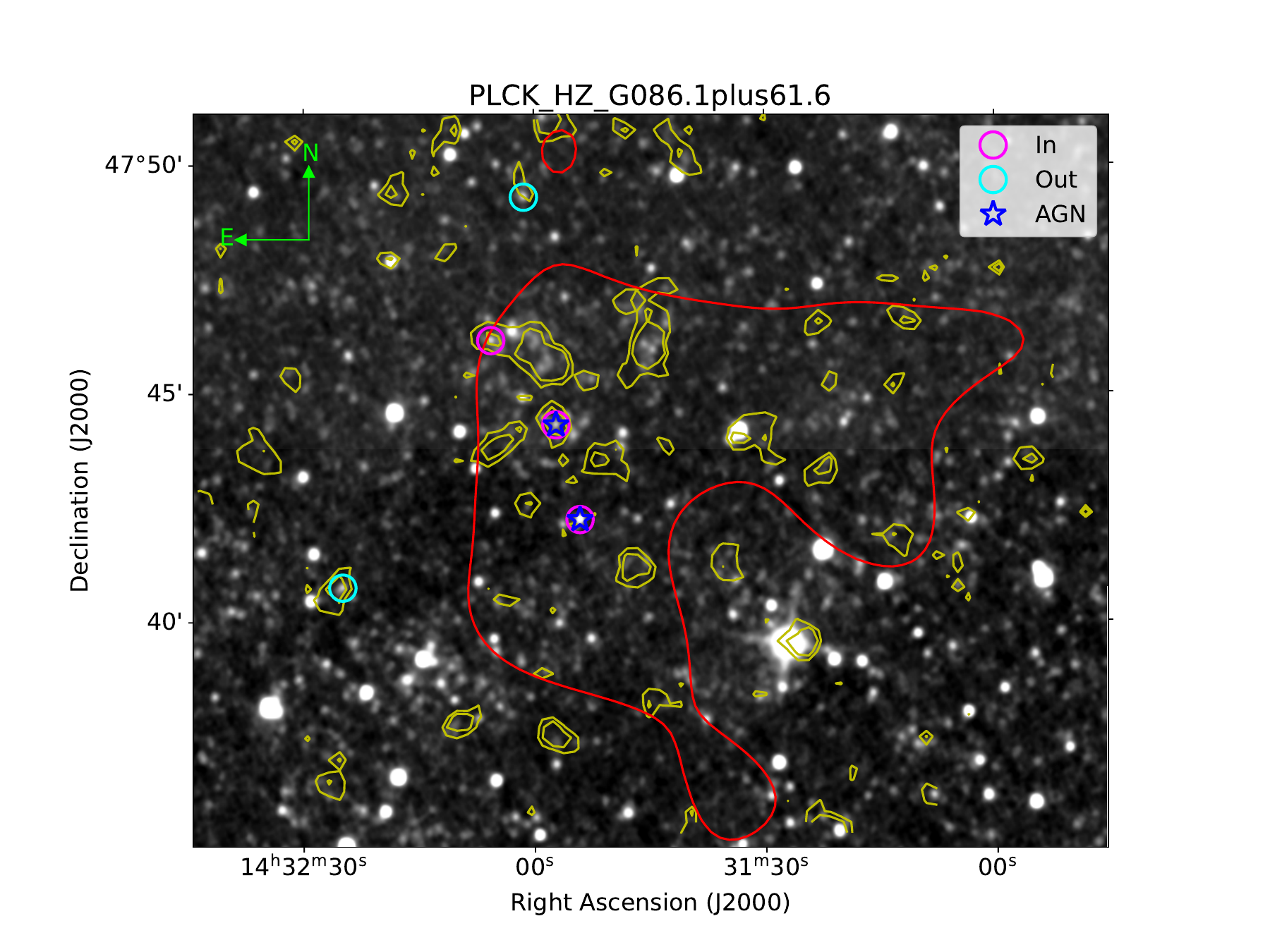}
    \caption{A $20 \times 16$ arcmin$^2$ WISE observation at $4.6 \mu m$ (W2 band) of one of our PC candidates, PLCK\_HZ\_G086.1plus61.6, shown as an example. The yellow contours show emission levels at 2$\sigma$ and 3$\sigma$ of the Herschel/SPIRE observation at $500 \mu m$, for the same field. The red contour corresponds to the $50\%$ of the peak flux of the respective Planck image at 545 GHz, which separates the `in' and `out' regions. WISE `in' and `out' sources are enclosed by magenta and cyan circles, respectively. The sources enclosed by a blue star were classified as AGN according to our method (see Section \ref{sec:agn-class}).}
    \label{PCexample}
\end{figure}

\subsection{Main sample}
\subsubsection{Planck XXVII} \label{sec:planck27}

Our main sample consists of the Herschel/SPIRE follow-up observations of 228 Planck sources from \cite{PlanckCollaboration2015}. These fields, selected as cold sources of the cosmic infrared background (CIB) and from the Planck catalogue of Compact Sources (PCCS), were chosen for follow-up because their rest-frame far-infrared colours show a peak between the frequency range 353-857 GHz, allowing the selection of ultra luminous infrared galaxies.

This sample is dominated by dusty far-infrared galaxies, with high star formation rates, suggesting the signatures of highly star-forming protoclusters at high redshift, some line-of-sight projections \citep{Negrello2017}, and strongly-lensed sources. Therefore, it is important to note that this study targets a specific population of galaxies in protoclusters, i.e. their most star-forming population.

Particularly for this study, we have discarded the Herschel/SPIRE sources that are considered lensed \citep[Dole H., private communication]{Canameras15}. After removing the lensed sources, we are left with 193 Planck sources.

Although this catalogue offers a good opportunity to study a large number of star-forming galaxies in protoclusters, it does not provide certain redshift measurements nor does it have enough multiwavelength observations to derive a redshift estimation such as photometric redshifts. However, we do have an idea of the redshift range for these sources.

First, since these sources are considered  `cold' sources of the cosmic infrared background (CIB), we know that they are at redshifts $z>1$ because the CIB is considered a proxy of intense star formation as those redshifts \citep[and references therein]{PlanckCollaboration2015, Planck2014}.    

More specifically, Planck observations show that these sources have spectral energy distributions (SEDs) that peak around 353 and 857 GHz, which equates to redshifted infrared galaxies at $z \sim 2-4$ \citep{PlanckCollaboration2015}. 

Also, \cite{PlanckCollaboration2015} followed the approach of \cite{Amblard2010}, and found a suggested redshift range of $z \sim 1.5-3$ with their Herschel colours and SEDs of modified blackbodies, with the redshift distribution of the SPIRE sources peaking at $z=$2 or 1.3 for dust temperatures of $T_{\mathrm d}$ = 35K or 25K, respectively.

Lastly, some protoclusters have been confirmed from this sample, at redshifts $z\sim 1.1-3.3$ \citep{Berman2022}, $z\sim 1.3-3$ \citep{Polletta2022}, $z\sim 1.5$ and 2.4 \citep{Kneissl2019}, $z\sim 1.7-2.0$ \citep{Flores-Cacho2016}, $z\sim 2$ \citep{Lammers2022} and $z\sim 2.16$ \citep{Koyama2021,Polletta2021}; and were also followed up by Spitzer at $z\sim 1.3-3$ \citep{martinache18}. This suggests a redshift range of $z\sim 1-3$.

\begin{figure}
    \centering
    \includegraphics[width=\columnwidth]{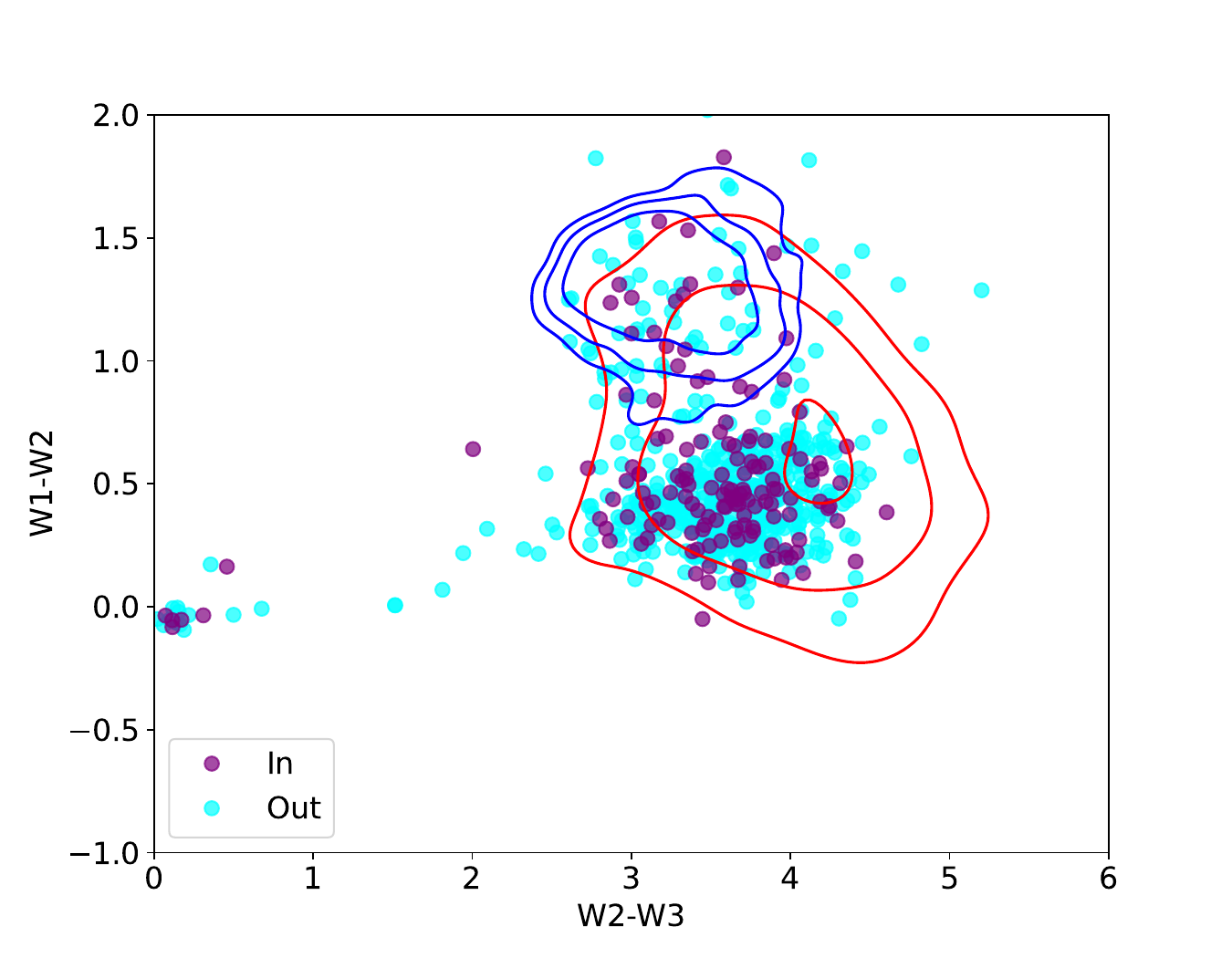}
    \caption{W1-W2 vs. W2-W3 colour-colour diagram of the WISE counterparts for the SPIRE sources. Purple and cyan dots show the sources that are `in' and `out' of the Planck 50\% intensity region, respectively. The `in' displays the same spread in colours as the sources in the `out' region. The over-plotted contours show the colour distribution for our control sample at the 1, 1.5 and 2$\sigma$ levels. The blue contours show the distribution for AGN sources, while the red contours show the non-AGN sources.
    Our control sample is used to train and test our classifier (see Section \ref{sec:agn-class}) and includes the sources described in Table \ref{table1} (see Section \ref{sec:controlsample}). We see that most of the SPIRE sources have colours in the same range as the colours from the control sample. The AGN distribution tends to be redder in the W1-W2 colour and bluer in the W2-W3 colour when compared to non-AGN galaxies.}
    \label{fig:dataSPcolors}
\end{figure}

The photometric catalogue for this sample includes the Herschel/SPIRE photometry for a total of 7099 for the 228 fields, or 6904, for the 193 fields left after removing lensed sources. The Herschel/SPIRE photometry has angular resolutions (FWHM) of $18\farcs1$, $25\farcs2$ and $36\farcs6$ at $250 \mu$m, $350 \mu$m and $500 \mu$m respectively \citep[]{griffin10}. For reference, Planck observations have angular resolutions (FWHM) of $4.6\arcmin$, $4.8\arcmin$ and $4.9\arcmin$ for the 857, 545 and 353 GHz maps, respectively \citep[Planck XXXIX]{PlanckXXXIX}.

\cite{PlanckCollaboration2015} separated the SPIRE sources in two regions, the `in' region and the `out' region. The `in' region is defined as the $50\%$ Planck intensity contour at 545 GHz, the map with the best signal-to-noise ratio (SNR), and has an approximate radius of $\sim 5$ arcmin. \cite{PlanckCollaboration2015} did an statistical analysis on the number counts for these regions and compared them with two control samples, the HerMES `level 5' Lockman-SWIRE field \citep{hermes10} and the Herschel Lens Survey (HLS) cluster fields at $z<1$ of \cite{Egami2010}. The statistical analysis shows that IN regions exhibit a chromatic excess consistent with a population of high-redshift ($z = 2-4$) lensed candidates, IN regions in both 350$\mu m$ and 500$\mu m$ have higher counts when compared to samples of the Lockman field and the $z<1$ HLS cluster fields, IN regions have an excess of SPIRE sources, and that OUT regions have number counts consistent with the Lockman field and the HLS cluster fields with similar density to blind surveys. Thus, the analysis suggests that the IN and OUT regions would be a good method for selecting PC member candidates and field sources, respectively. For instance, this approach is used by \cite{Lammers2022}.

In Figure \ref{PCexample} we show the WISE image of one of the PC candidates as an example. We show the W2 band image for the field PLCK\_HZ\_G086.1plus61.6, along with the `in'  and `out' sources (pink and cyan circles, respectively). The Herschel 500$\mu$m emission is also shown in yellow contours, showing the difference in resolution between Herschel and WISE. Also, we show the contour at 50\% of the peak flux for the Planck image at 545 GHz, which separates the `in'  and `out' regions.

\subsubsection{Planck XXVII WISE counterparts}
\label{sec:wise-count}

The WISE survey is an all-sky mid-IR survey at 3.4 $\mu m$, 4.6 $\mu m$, 12 $\mu m$ and 22 $\mu m$ (the W1, W2, W3 and W4 bands, respectively) with angular resolutions of $6\farcs1$, $6\farcs4$, $6\farcs5$ and $12\farcs0$. WISE data provide observations with higher resolution compared to Herschel/SPIRE. Also, while Herschel probes dust heated by ongoing star formation, WISE W1 and W2 bands probe the stellar emission. Furthermore, WISE observations have been demonstrated to be especially efficient at revealing the presence of an AGN, due to the mid-IR emission of AGN-heated dust \citep[]{stern12, mateos12, assef13, hviding22}.

We have thus decided to take advantage of the AGN diagnostic power provided by WISE, to assess the presence of AGN activity in the Planck XXVII PC sample, by using the WISE counterparts of our SPIRE sources. Moreover, since we expect PC members to be bright and red sub-mm sources, we are reducing contamination from non-members by only selecting WISE sources associated with Herschel sources. 

The SPIRE sources were cross-matched with the AllWISE data release\footnote{Explanatory supplement to the AllWISE data release products in \url{https://wise2.ipac.caltech.edu/docs/release/allwise/}} \citep[]{wise10, neoWISE}, using the public database from the NASA/IPAC Infrared Science Archive \footnote{\url{https://irsa.ipac.caltech.edu/}} (IRSA). The match was done with the SPIRE 250 $\mu m$ band and considered only the closest source as a counterpart (avoiding multiple counterparts) in a radius of 9\arcsec, which is half the resolution of the SPIRE's 250 $\mu m$ band. This was a conservative choice to limit the wrong associations. We also considered that the WISE sources were photometrically not affected by contamination or artefacts (\verb|cc_flags=`0000'|) and that they were point-like (\verb|ext_flg=0|) as expected for high-redshift sources. We use \verb|w#mpro| Vega magnitudes (where \verb|#| in the observing band 1, 2, 3 or 4), which is the appropriate magnitude of non-extended sources.

\begin{figure*}
    \centering
    \includegraphics[width = \textwidth]{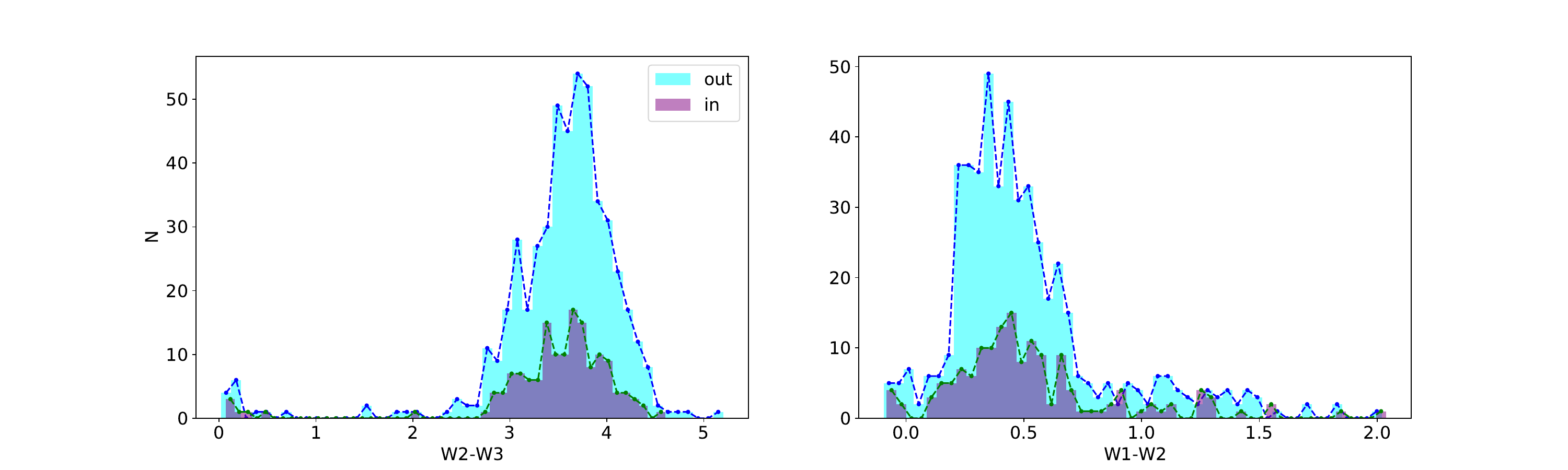}
    \caption{SPIRE WISE counterparts distributions for the W2-W3 (left panel) and W1-W2 (right panel) colours. Purple and cyan colours represent sources flagged as \textit{in} and \textit{out}, respectively. The green and blue dashed curves show the interpolation to the \textit{in} and \textit{out} distributions. We used these interpolations to generate the simulated data for the Monte Carlo simulation (used to estimate the uncertainty of the classifier).}
    \label{fig:sphericDists}
\end{figure*}

After the cross-match, it was necessary to choose a SNR threshold for the WISE bands to obtain a trustworthy sample of counterparts. To decide on an SNR threshold, we derived the completeness level of the sample at different SNR values.
For the estimates of the completeness of the sample, we searched all sources in AllWISE within a search radius of $20$ arcmin from the Planck's field centres, which is a few times larger than a typical Planck `in' region, obtaining more than 1 million sources. Then, we compared the mean fluxes of the sample with different SNR limits for each band, with the AllWISE catalogue completeness. Details on how this completeness is computed are found in the AllWISE documentation  \footnote{ \url{https://wise2.ipac.caltech.edu/docs/release/allwise/expsup/sec2_4a.html} for W1 and W2 and \url{https://wise2.ipac.caltech.edu/docs/release/allsky/expsup/sec6_5.html} for W3 and W4.}.
After this comparison, we decide on an SNR threshold of SNR $ \geq 7$, which corresponds to a completeness level of at least $\sim$ $45 \%$ for our AllWISE counterparts sample. We argue that a higher completeness level is not required, considering that our sources from the Planck XXVII catalogue are secure SPIRE detections. Also, this SNR threshold is consistent with a confident point source detection \citep{lonsdale15}. 

Finally, from our 6904 SPIRE sources, we obtained 646 AllWISE counterparts. Out of the total AllWISE sample, 150 are considered PC members (or `in' sources) and 496 are considered field galaxies (or `out' sources).
A WISE W1-W2 vs. W2-W3 colour-colour diagram of our AllWISE sources is shown in Figure \ref{fig:dataSPcolors}. Here, we show both the PC members (`in') and field galaxies (`out') sources.

\begin{table*}
    \caption{Summary of our final control sample. For each catalogue, we show the type of galaxy selected, the number of sources and the corresponding reference.}
    \label{table1}
    \centering
    \begin{tabular}{lccl}
    \hline
Catalogue name & Type & N selected sources & Reference \\  \hline
Million Quasars (Milliquas) v7.2 & AGN & 619 & \cite{Flesch2021} \\
AGNs in the MIR using AllWISE data & AGN & 8 & \cite{Secrest} \\
VERONCAT & AGN & 9 & \cite{VC10} \\
FMOS-COSMOS & non-AGN & 516 & \cite{Kashino} \\ 
CANDELS-EGS     & non-AGN & 64  & \cite{Stefanon2017} \\ 
CANDELS-COSMOS  & non-AGN & 10  & \cite{Nayyeri2017} \\
CANDELS-UDS     & non-AGN &  26 & \cite{Santini2015} \\
CANDELS-GOODS-S & non-AGN &  20 & \cite{Santini2015} \\
\hline 
Total   &   & 1272 & \\ \hline
\end{tabular}
\end{table*}

\subsection{Control Sample}
\label{sec:controlsample}

To train the AGN classifier, a control sample was compiled with a combination of catalogues for AGN and non-AGN sources with available WISE colours. Considering the suggested redshift ranges for the SPIRE sources (discussed in Section \ref{sec:planck27}), and the redshift range of the confirmed structures from the sample, sources were selected between $1\leq z\leq3$.

If the catalogue includes the WISE photometry, then the magnitudes and colours were retrieved from the catalogue itself. Otherwise, a cross-match with AllWISE was done, following the same procedure of the SPIRE sources, but using a search radius of $6 \arcsec$, which corresponds to half the best angular resolution of the WISE bands.

\subsubsection{AGN sources} \label{sec:AGN}

For the AGN subsample, we selected AGN sources from the Million Quasars (Milliquas) catalogue, version 7.2  \citep{Flesch2021}, the AGNs in the MIR using AllWISE data \citep{Secrest} catalogue,  and the Veron catalogue of Quasars \& AGN, 13th Edition (VERONCAT; \citealt{VC10}).

The \cite{Flesch2021} catalogue corresponds to a compilation of $\sim 800,000$ quasars up to 30 April 2021, and is the updated version of the \cite{Flesch2015} catalogue. It includes different types of sources, and we only selected secure quasar objects.

The \cite{Secrest} catalogue is an all-sky sample that contains more than 1 million sources ($>66,000$ within $1\leq \z \leq3$), including previously uncatalogued AGN from the AllWISE data release \citep[]{wise10, neoWISE}, using two-colour infrared selection criteria.

The VERONCAT catalogue contains 168,941 objects of different types of AGN (99,848 sources within $1\leq \z \leq 3$). This catalogue is a compilation of all AGN known from the literature (until the VERONCAT's publication date), including data from the 2dF \citep[]{croom01, croom04} catalogue and the data releases (from 1st to 7th; \citealt{sdss1, sdss2, sdss3, sdss4, sdss5, sdss6, sdss7}) of the Sloan Digital Sky Survey (SDSS; \citealt{SDSS}) catalogue.

\subsubsection{Non-AGN galaxies} \label{sec:noAGN}

Non-AGN sources were selected from the catalogue of star-forming galaxies at $z \sim 1.6$ in the FMOS-COSMOS survey from \cite{Kashino} and the catalogues from the Cosmic Assembly Near-IR Deep Extragalactic Legacy Survey  (CANDELS \footnote{ \url{https://archive.stsci.edu/hlsp/candels}};  \citealt{Grogin2011,Koekemoer2011}). Particularly, we used the GOODS-S CANDELS and UDS CANDELS stellar mass catalogues from \cite{Santini2015}, the CANDELS-EGS stellar mass catalogue from \cite{Stefanon2017} and the 
CANDELS-COSMOS Multiwavelength catalogue from \cite{Nayyeri2017}.

The catalogue from \cite{Kashino} contains 5,484 objects observed over the COSMOS field, with $\sim30 \%$ of them being within $1 \leq \z \leq 3$. AGN sources were discarded using catalogues of X-ray sources \citep{Kashino}.
Only 516 sources remained, after cross-matching, with AllWISE photometry.

The CANDELS catalogues \citep{Stefanon2017,Nayyeri2017,Santini2015} were chosen because they include an AGNflag, which allows the selection of non-AGN sources. This flag comes from SED fitting of multi-wavelength observations.

\subsubsection{Balanced control sample}
\label{sec:balancedControlSample}
The resulting count of sources is a total of $\sim 21,000$ AGN sources and 636 non-AGN.
The number of AGN sources is much greater than the number of non-AGN sources, but a well statistically balanced sample is necessary to train the classifier, to avoid biases.

Therefore, we reduced our AGN sample to match the number of non-AGN sources. For this, we randomly selected 636 AGN sources. Then, we have 636 AGN and 636 non-AGN sources, and a total of 1272 sources, summarised in Table \ref{table1}.
 
The colour-colour diagram for our final control sample can be seen in the right panel of Figure \ref{fig:dataSPcolors}. Here we distinguish between the AGN and non-AGN samples. At first glance, no strong separation between AGN and non-AGN can be seen. However, the AGN distribution tends to be redder in the W1-W2 colour and bluer in the W2-W3 colour when compared to non-AGN galaxies.

\begin{figure*}
\includegraphics[width = \textwidth]{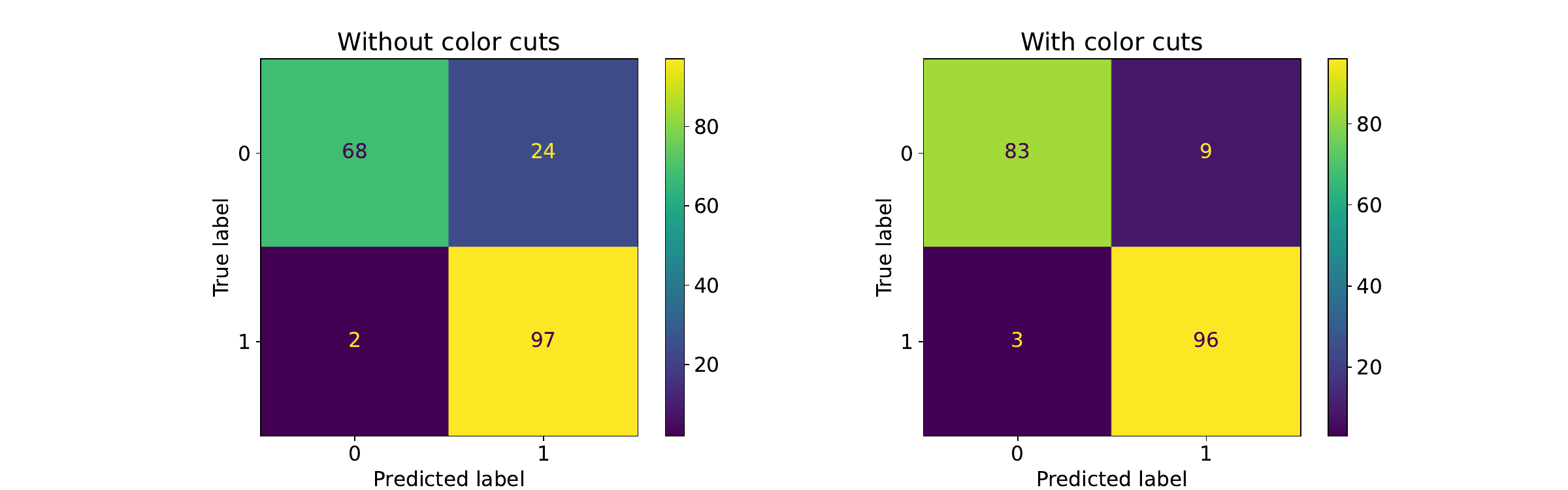}
        \caption{Confusion matrices of our classification test with a 191 sources test sample. Each matrix shows the number of true positives (bottom right), false negatives (bottom left), false positives (top right) and true negatives (top left). \textit{Left panel:} Confusion matrix for the classifier without adding the colour cuts at W1-W2 > 0.94 and W2-W3 < 4.04. The accuracy of the classification is 86\%, with 98\% completeness and 80\% reliability. \textit{Right panel:} Confusion Matrix for the classifier, now adding the colour cuts at W1-W2 > 0.94 and W2-W3 < 4.04. The accuracy of the classification using the colour cut increases to 94\%, with 97\% completeness and 91\% reliability.}
    \label{fig:testing}
\end{figure*}

\begin{figure}
    \centering
    \includegraphics[width = 0.5\textwidth]{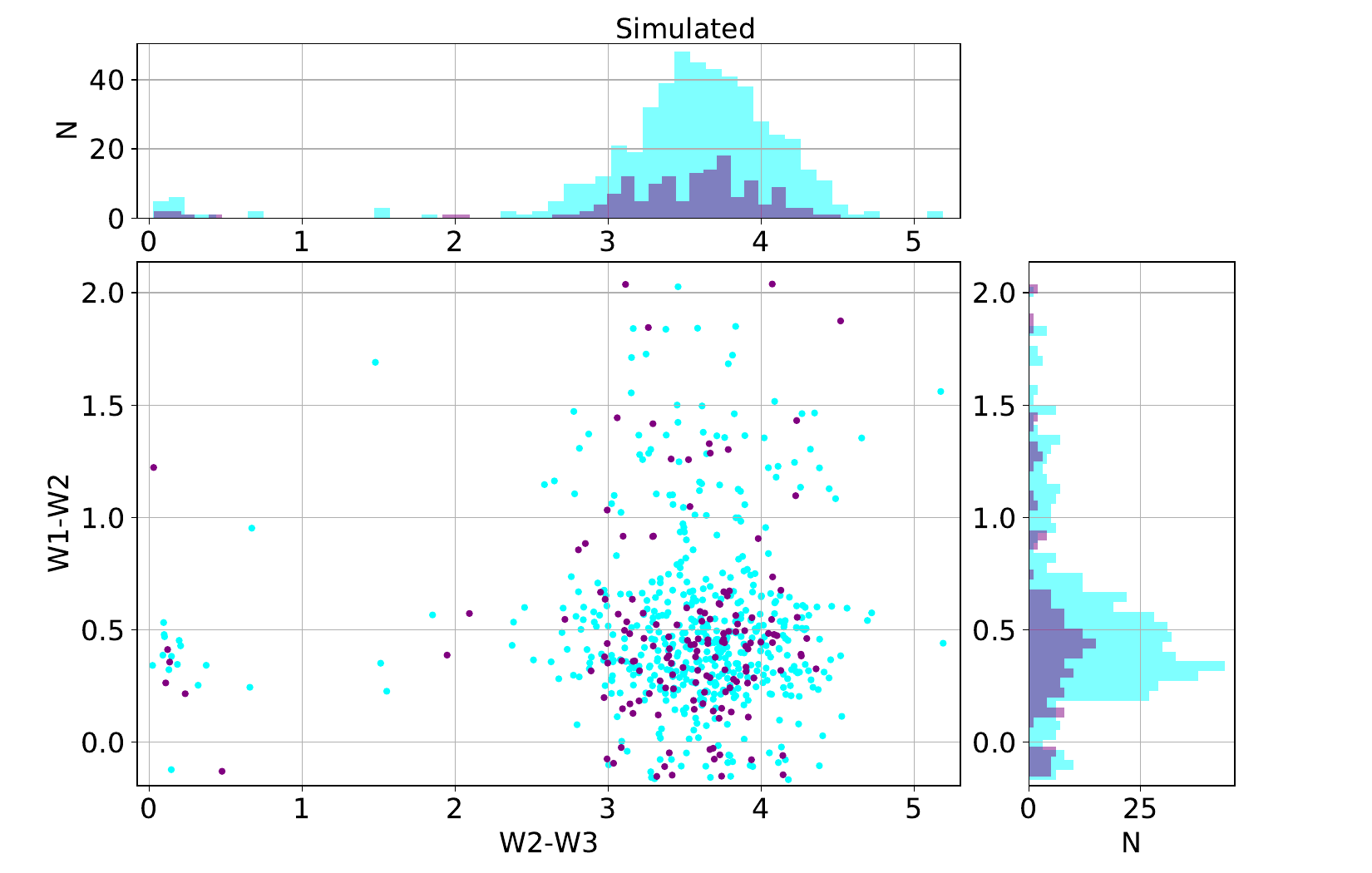}
    \caption{Example of simulated data for one iteration of the Monte Carlo simulation. The figure shows the colour-colour diagram for W1-W2 as a function of W2-W3. Purple and cyan colours represent sources flagged as \textit{in} and \textit{out}, respectively. Here we show that the simulated data follow a similar colour-colour distribution to the SPIRE's one, shown in Figures \ref{fig:dataSPcolors} and \ref{fig:sphericDists}.}
    \label{fig:simDataSP_dists_colors}
\end{figure}

\section{AGN classification}
\label{sec:agn-class}

We design a colour-colour selection criterion (i.e. a classifier) to sort galaxies of our SPIRE sources as AGN or non-AGN galaxies. This is achieved by finding a way of separating both types of galaxies on a W1-W2-W3 colour-colour space. This classifier uses two main criteria. First, a \textit{K-means} clustering machine learning algorithm \citep{Macqueen67,Lloyd82} is applied to the WISE colour-colour diagram of known AGN and non-AGN sources (i.e. the control sample). After that, a mid-IR/WISE colour selection criterion is applied. In this case, we use two colour cuts of W1-W2 and W2-W3 (see next subsection). The colour cut at W1-W2 > 0.94 is higher than the value used in other studies (e.g. 0.8 in \citealt{stern12} and 0.5 in \citealt{blecha18}).

This type of classifier is based on similar studies, that were able to distinguish different types of galaxies, mostly at lower redshifts, in a colour-colour diagram of WISE W1-W2 and W2-W3 colours \citep{Lake2012,Mingo2016,Jarrett2017}.

After classifying the WISE counterparts of our SPIRE sources, we
estimate the AGN fractions for both the PC members (`in' sources)
and field galaxies (`out' sources). The AGN fraction uncertainty is
estimated via a Monte Carlo approach.

\subsection{Building the classifier}

Before training our classifier, we subdivide our control sample into a training set, that corresponds to an $85\%$ of the full sample, and a test set, corresponding to the $15\%$ left of the sample. This resulted in 1,081 galaxies for the training set and 191 galaxies for the test set. The percentages that we used to make each sub-sample were decided based on having enough sources to train the classifier, and enough sources that allowed us to evaluate the accuracy of the classifier.

The first part of the classifier consists of using a \textit{k-means} algorithm from the \textsc{Python} package \textsc{Scikit-learn} \citep{Pedregosa11}. \textit{K-means} is an unsupervised, machine-learning, clustering algorithm. This algorithm subdivides the sample into clusters so that the sum of the squares of the data values in the W2-W3 vs W1-W2 colour-colour space within each cluster is minimised.

The \textit{K-means} module uses the \textit{K-elbow} parameter to decide how many clusters the algorithm will divide the data into. Considering that we want to distinguish between AGN and non-AGN, we set the \textit{K} parameter as $K = 2$, thus dividing the data into two clusters.

\begin{figure*}
    \centering
    \includegraphics[width = \textwidth]{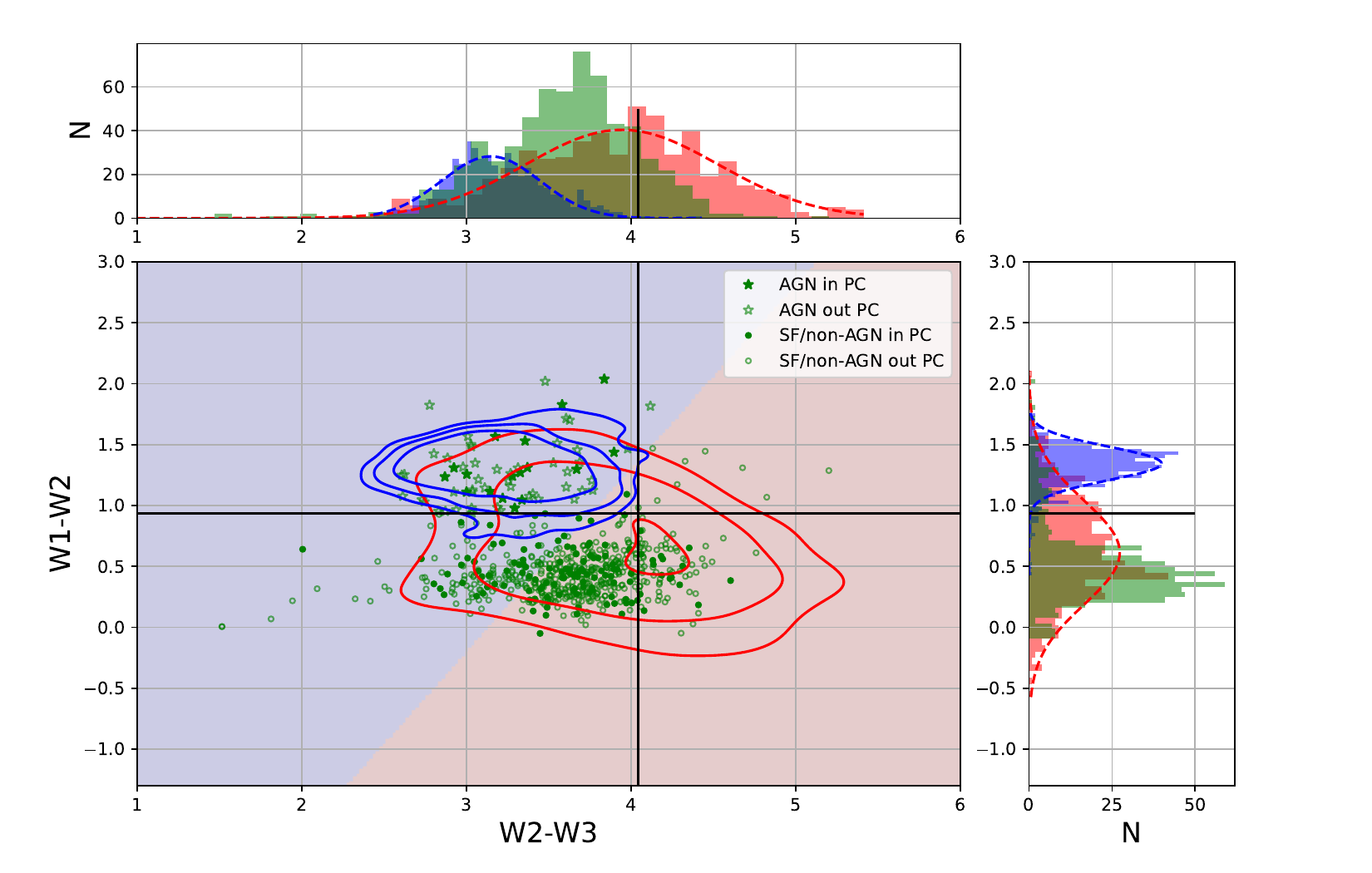}
    \caption{Classification result for the SPIRE sources (green) in the W1-W2 vs. W2-W3 colour-colour diagram. To consider a source as an AGN, three conditions must be met. The source must be located: (1) above the black horizontal line, which corresponds to a 3$\sigma$ level threshold for AGN in W1-W2 colour, (2) to the left of the black vertical line, which is the 3$\sigma$ level threshold for AGN in W2-W3 colour, and (3) over the red background area of the colour-colour diagram, which corresponds to the AGN classification given by the \textit{k-means} separation. Filled (empty) stars represent the sources inside (outside) the PCs that were classified as AGN. Filled (empty) circles are the sources inside (outside) the PCs classified as non-AGN. The blue and red contours show the 1, 1.5 and 2$\sigma$ levels of the AGN and SF/non-AGN sources in the training data set of our control sample, respectively. This shows that both our main sample and control sample have a similar colour range covered. The upper panel shows the histogram distribution for the W2-W3 colour  and the 3$\sigma$ level threshold shown as the solid black line for the SPIRE sources (green), and for the AGN sources (blue) and the non-AGN sources (red) of the control sample. Similarly, the right side panel shows the histogram of those distributions for the W1-W2 colour,  including the 3$\sigma$ level threshold shown as the solid black line. The different coloured dashed lines show the fitted Gaussian model for each distribution.}
    \label{fig:kmeansV1}
\end{figure*}

Once the algorithm finishes assigning every data point in the training set to a given cluster, each point gets flagged with either 1 or 0, which means that the source was selected as either an AGN or a non-AGN, respectively. The separation is given by W1-W2 = 1.53(W2-W3) - 4.80. Since running the \textit{k-means} algorithm alone does not cleanly divide our sample, we added two colour cuts into the classifier. The colour cuts were defined as the mean minus 3$\sigma$ of the W1-W2 AGN distribution, and as the mean plus 3$\sigma$ of the W2-W3 colour from our control sample. This corresponds to colour cuts at W1-W2 > 0.94 and W2-W3 < 4.04 (see Figure \ref{fig:kmeansV1}). In summary, we consider a source to be an AGN if its W1-W2 and W2-W3 colours agree with the following:

\begin{equation}
    W1-W2 > \begin{cases}
    0.94 & \mathrm{,\ } W2-W3  \leq 3.76 \\
    1.53(W2-W3) - 4.80 & \mathrm{,\ } 3.76 < W2-W3 < 4.04
    \end{cases}
    \label{eq:color-criteria}
\end{equation}

\subsection{Testing} \label{sec:testing}

We estimate the completeness, reliability and accuracy of the classifier using the test sample, with a size of 191 sources. We first verify the classification only by considering a \textit{k-means} clustering. This results in a classification of 97 true positives and 68 true negatives. Considering the completeness as the number of true positives divided by the sum of true positives and false negatives, we get a completeness of $98\%$. For the reliability, measured as the number of true positives divided by the sum of true positives and false positives, we get a reliability of $80\%$. Lastly, for the accuracy, measured as the sum of true positives and true negatives divided by the total number of sources, we get an accuracy of $86\%$.

 We then tested these parameters using the combined \textit{k-means} algorithm with the colour cut criterion. This resulted in 83 true negatives and 96 true positives. This essentially means that adding the colour cut improves the accuracy of our classifier to a $94\%$, with a $97\%$ completeness and a $91\%$ reliability. In Figure \ref{fig:testing} we present the confusion matrices, summarising these values.

\subsection{Monte Carlo simulation}
To estimate the uncertainty of our method we performed a 10,000-step Monte Carlo simulation. Each step simulates colours in the space W1-W2 vs. W2-W3, for which we use our classifier and then measure an AGN fraction. 
To construct our simulation, we interpolate the W1-W2 and W2-W3 colour distributions of the SPIRE sources.
The SPIRE distributions in each colour and their interpolations are shown in Figure \ref{fig:sphericDists}.

Each step simulates the data using random data points generated from the previously mentioned distributions. The W1-W2-W3 colours of one of the artificially generated distributions are shown in Figure \ref{fig:simDataSP_dists_colors}. After that, each simulated data point gets a designation of AGN or non-AGN, using our classifier. Finally, the AGN fractions are measured. The Monte Carlo simulation returns a normal distribution, in which the standard deviation $\sigma$ is the corresponding uncertainty.

\section{Results}
\label{sec:results}

\subsection{Classification of the SPIRE sources}

We ran the classifier with our SPIRE sources and found the following outcome. Out of the full catalogue of 646 sources, we found that 64 were selected as AGN, while 582 were selected as non-AGN sources. In particular, we found that there are 17 AGN that correspond to members of PC candidates and 47 that correspond to sources outside the PC candidates, i.e. field galaxies. When it comes to non-AGN, we found 133 that are also PC members and 449 sources that correspond to non-PC members. These numbers are summarised in Table \ref{tab:classificationResult}.

\begin{table}
    \caption{Classification result of the SPIRE WISE counterparts.}
    \label{tab:classificationResult}
    \centering
    \begin{tabular}{lcc}
    \hline
         & AGN & non-AGN \\
    \hline
        In & 17 & 133 \\
        Out & 47 & 449 \\
        Total & 64 & 582 \\
    \hline
    \end{tabular}
\end{table}

The final classifier is represented in Figure \ref{fig:kmeansV1}. In particular, we show the W1-W2 vs W2-W3 colour-colour diagram for our SPIRE sources. The different symbols distinguish between the sources that the classifier selects as AGN or non-AGN. Also, filled and empty symbols differentiate member galaxies of PC from field galaxies, respectively. We also over-plotted the (training) control sample as blue and red contours for AGN and non-AGN objects, respectively, to show the distribution of the galaxies we used to train our \textit{k-means} method.

\subsection{AGN fractions}

After the classification of sources in our SPIRE sample, we proceeded to measure the AGN fraction in both the PC candidates and the field. The resulting AGN fraction for PCs is $f_{AGN_{in}} = 0.113 \pm 0.03$ or $11\pm3\%$. For the field, we found an AGN fraction of $f_{AGN_{out}} = 0.095 \pm 0.013$ or $10\pm1\%$.

The uncertainties to each AGN fraction come from Monte Carlo simulations. The Monte Carlo histograms are shown in the top panels of Figure \ref{fig:arrs}, of the Appendix \ref{sect:monte-carlo}. We note that the AGN fractions measured by the Monte Carlo simulations are quite similar to the actual AGN fractions. This is a good probe that our simulated data are a good representation of the observations.

For a better understanding of our results, we also measured the AGN fraction for `red' SPIRE sources. The `red' sources come from the selection of the reddest Herschel sources by \cite{PlanckCollaboration2015}, defined as $S_{350}/S_{250} > 0.7$ and $S_{500}/S_{350} > 0.6$, based on source density distributions. This sample of SPIRE red sources has a higher overdensity significance than the SPIRE sources \citep[see Figures 6 and 7]{PlanckCollaboration2015}, suggesting this method as another way of selecting PC members.
Therefore, in this case, we measure AGN fractions for PC members and non-members, considering red SPIRE sources as the PC member candidates and the non-red sources as field galaxy sources.

The AGN fraction of red SPIRE sources is $f_{AGN_{red}} = 0.186 \pm 0.044$ or $19\pm4\%$. For the `non-red' sources, the AGN fraction is $f_{AGN_{non-red}} = 0.037 \pm 0.010$ or $4\pm1\%$. The Monte Carlo histograms showing the estimated uncertainty are in the middle panels of Figure \ref{fig:arrs}, of the Appendix \ref{sect:monte-carlo}.

At first glance, if we consider the PC members as `in' sources and field galaxies as `out' sources, we find an AGN fraction higher in PC candidates, but with a difference not statistically significant considering the uncertainties of our estimates. However, if we consider the PC members as the `red' sources and field galaxies as the `non-red' sources, we find a clear increase of AGN fraction in the PC candidates with respect to the field, by at least a factor 3 (with 1$\sigma$ uncertainty).

\begin{figure}
    \centering
    \includegraphics[width = \columnwidth]{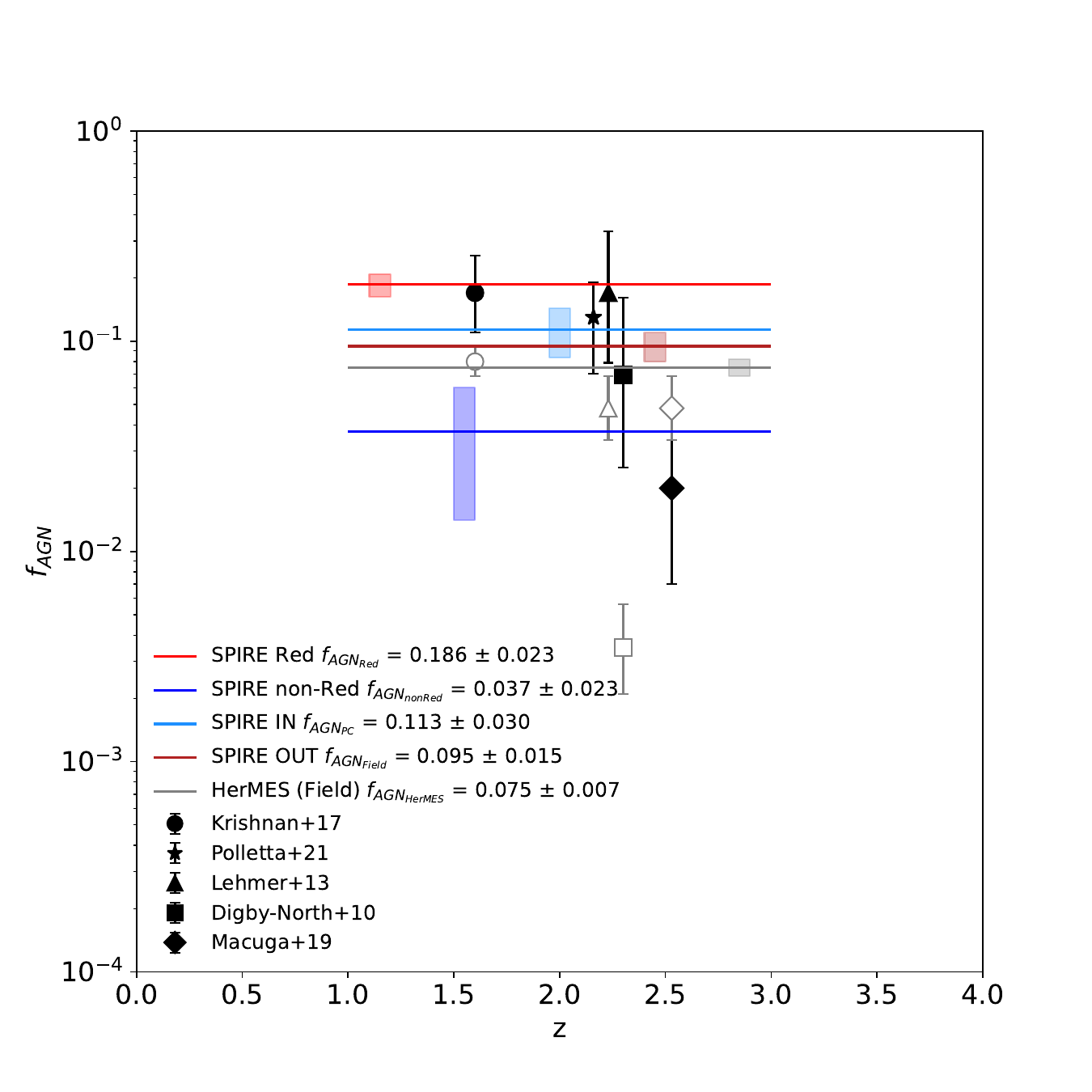}
    \caption{AGN fractions, $f_{AGN}$, for the red (blue), non-red (red), `in' (cyan) and `out' (dark red) SPIRE sources, and HerMES field (grey) versus redshift. For easier visualisation, we show the 1$\sigma$ significance of the AGN fractions as boxes in arbitrary redshift positions. Literature values from \protect\cite{Polletta2021} (black star) and \protect\citet[and references therein; black circle, triangle, square and diamond]{Macuga19} are added as reference for cluster/PC (filled black markers) and field (empty grey markers) galaxies. Here we show that the AGN fraction is, in general, greater in PCs than in the field. See also Figure \ref{fig:AGNFractionResult2}.}
    \label{fig:AGNFractionResult}
\end{figure}

\begin{figure}
    \centering
    \includegraphics[width = \columnwidth]{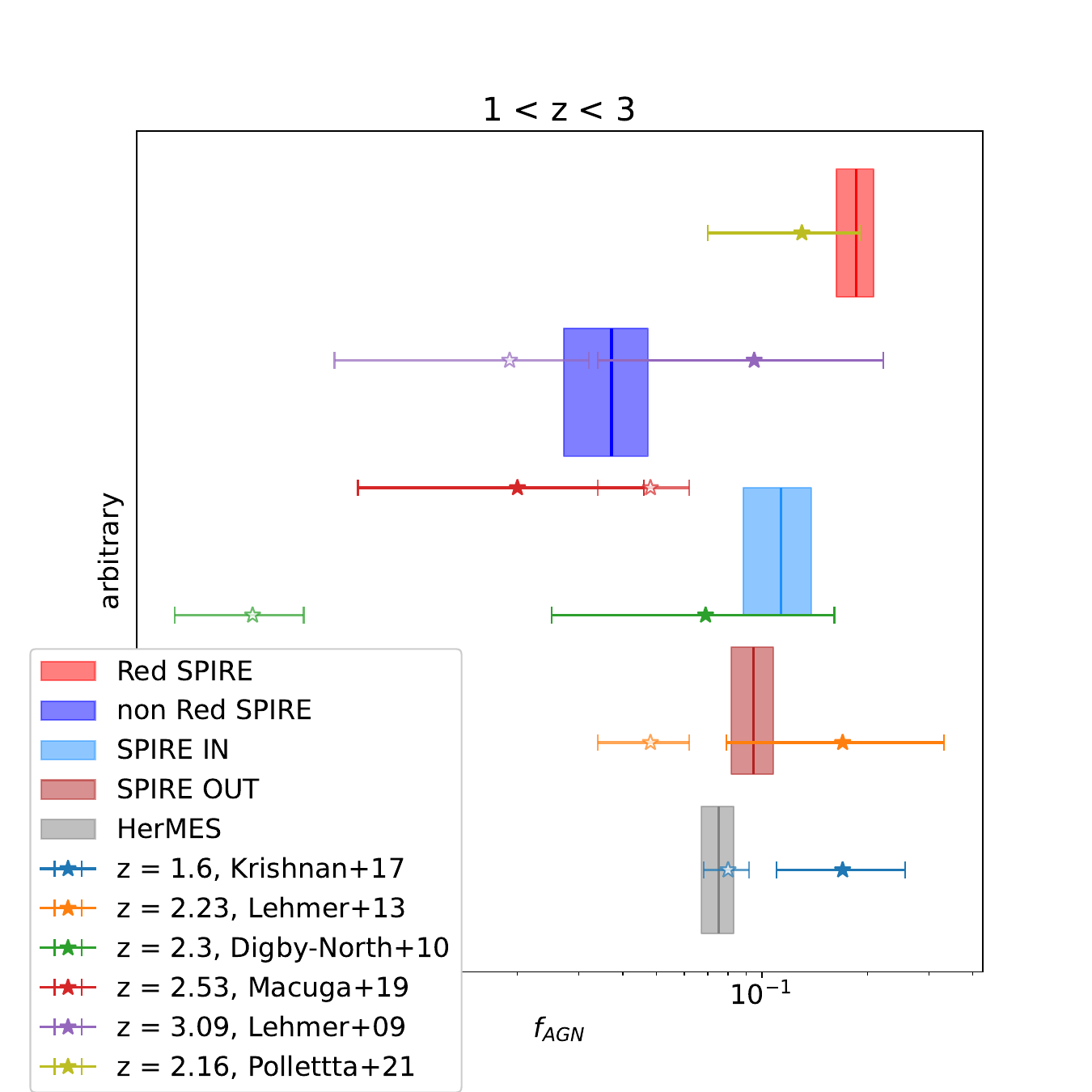}
    \caption{AGN fractions, $f_{AGN}$ (and 1$\sigma$ significance), for the red SPIRE sources (blue), non-red SPIRE sources (red), SPIRE sources that are inside PC (light blue), outside PC (dark red) and HerMES field (grey). Literature values from \protect\cite{Polletta2021} and \protect\citet[and references therein]{Macuga19} are added as references for cluster/PC (filled stars) and field (empty stars) galaxies. The arbitrary y-axis was chosen to better distinguish the difference in the AGN fractions, taking into account their significance. See also Figure \ref{fig:AGNFractionResult}.}
    \label{fig:AGNFractionResult2}
\end{figure}

To compare these AGN fractions, we also measured the AGN fraction of the HerMES `level 5' Lockman-SWIRE field \citep[]{hermes10}, which has a similar depth to our SPIRE sources \citep{PlanckCollaboration2015}. We find that $f_{AGN_{HerMES}} = 0.075 \pm 0.008$ or $8\pm1\%$. The Monte Carlo histogram showing the estimated uncertainty is in the bottom panel of Figure \ref{fig:arrs}, of the Appendix \ref{sect:monte-carlo}. This AGN fraction is lower than the $f_{AGN_{out}}$. Figures \ref{fig:AGNFractionResult} and \ref{fig:AGNFractionResult2} summarise these fractions.

Since we do not have the exact redshift for each source and we are only working on a suggested redshift range of $1<z<3$, we plotted the fractions as a horizontal bar that extends through that redshift range. To compare our results, we added AGN fractions from \citet[and references therein]{Macuga19} at a redshift of z $= 2.53$. The figure also includes measurements for different PCs from \cite{Lehmer2009}, \cite{Digby2010}, \cite{Lehmer2013}, \cite{Polletta2021} and \cite{Krishnan2017}, at redshifts of z = 3.09, 2.3, 2.23, 2.16 and 1.6, respectively. We found similar values of $f_{AGN}$ in PCs to those in \cite{Krishnan2017} and \cite{Lehmer2013}, while the others seem lower. It is important to keep in mind that these studies only measured the AGN fraction within one PC, instead of a fraction within a large set of PC members, like in this study. It is also important to mention that these studies are based on different AGN selection approaches than this work, therefore it is difficult to compare them directly. However, they still mostly find an increasing number of AGNs in PCs than in the field.

\section{Discussion}
\label{sec:discussion}

\subsection{AGN selection}
We expect that training our AGN classifier with a richer data set would return a higher accuracy of classification and better statistical results, since here we were limited by a relatively small sample of star-forming galaxies at high redshift with WISE photometry. According to \citet{stern12} and references therein, one could decide on a different colour cut between the range $0.7 \leq W1-W2 \leq 0.8$, `trading' completeness (bluer colour cut) for reliability (redder colour cut), however, our W1-W2 colour cut is higher than this range (W1-W2 > 0.94).

\begin{figure*}
    \centering
    \includegraphics[width = \textwidth]{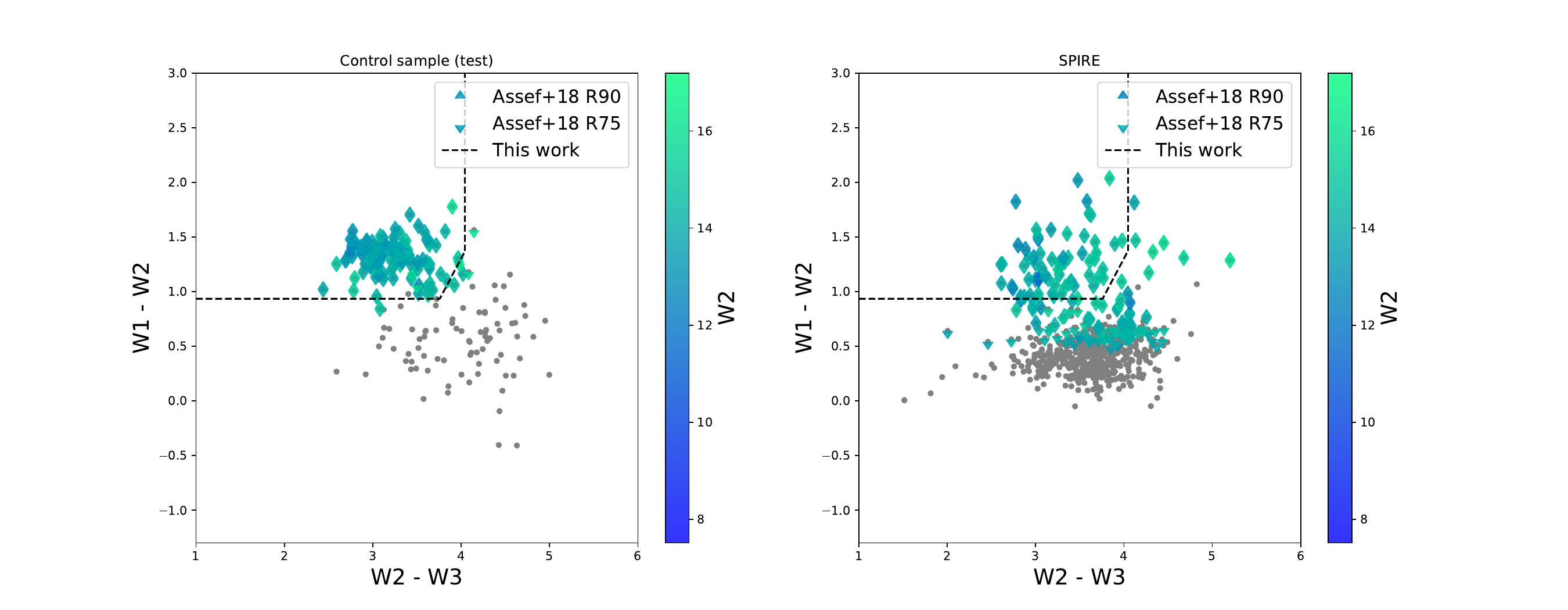}
    \caption{Comparison of AGN selection criteria between this work and \protect\cite{Assef18}. In the left (right) panel we show the colour-colour distribution of our control (SPIRE) sample (grey dots). The dashed black line shows our AGN selection criterion while up- and down-pointing triangles correspond to sources selected as AGN following \protect\cite{Assef18} for R90 and R75, respectively. The AGN selected sources are colour-coded for W2 magnitude. The majority of our AGN selected data, 109\% from the control sample, 84\% from the all SPIRE sample, and 95\% from the red SPIRE sources were also selected as AGN following \protect\cite{Assef18}'s criteria, principally for R90.}
    \label{fig:assefComp}
\end{figure*}

Keeping this in mind, plus the fact that our classifier shows an 94$\%$ of accuracy (see Section \ref{sec:testing} and Figure \ref{fig:testing}), we compared our classification method with the one shown in \cite{Assef18}, which also classifies AGN based on a colour condition. Particularly, we compared the number of SPIRE sources selected as AGN, following our criteria versus \cite{Assef18}.
This was made by comparing our equation \ref{eq:color-criteria} with equation 4 in \cite{Assef18}. The latter equation includes two outcomes, one for a reliability of 90\% and completeness 17\%, and the other for a reliability of 75\% and completeness 28\% (R90 and R75 respectively). 
We find that the ratio of AGN classified with our method and Assef's, for R90 (R75), is
$$\frac{AGN_{ThisWork}}{AGN_{Assef+18}} = 0.63 (0.38)$$ and
$$\frac{AGN_{ThisWork}}{AGN_{Assef+18}} = 0.78 (0.60)$$ for red SPIRE sources. Seeing this, the method from \cite{Assef18} tends to find more AGN than our method.
Doing this same analysis with our (test) control sample, we obtain that, $$\frac{AGN_{ThisWork}}{AGN_{Assef+18}} = 1 (0.95),$$ finding the same number of AGN than Assef's method with R90. However, when comparing the fraction of predicted AGN sources vs. true AGN sources ($True$) we find that: $$\frac{AGN_{Assef+18}}{True} = 1.06(1.11)$$ for R90(R75), and $$\frac{AGN_{ThisWork}}{True} = 1.06$$
This means that the criterion used in \cite{Assef18} can classify all AGN sources as an AGN, but includes an extra 6\% (11\%) of false positives for R90 (R75), while our method has an extra 6\% of false positives, slightly surpassing the $90\%$ reliability of Assef's method by $1\%$ and reaching a completeness of $97\%$.
In Figure \ref{fig:assefComp} we show a comparison between \cite{Assef18} and our AGN selection criteria. In Table \ref{tab:comparisonTable} we summarised this comparison.

We conclude that our method and the method from \cite{Assef18} are both useful and reliable methods to classify AGN. However, since the goal of this study is to measure AGN fractions, i.e. both the number of AGN and non-AGN are important, we argue that the most important characteristic of the classifier must be the completeness. In this case, our method is more appropriate because our completeness is $97\%$, while at the same time we reach a 91\% of reliability and an accuracy of 94\%, in contrast to the $17\%$ completeness of \cite{Assef18} for R90 or to the $28\%$ completeness for R75.

\begin{table}
\caption{Ratio of AGN classifications following this work and \protect\cite{Assef18} and the true number of AGN. General comparison between AGN classification from this work and classification from \protect\cite{Assef18}.}
\label{tab:comparisonTable}
\begin{tabular}{ccccccc}
\hline
\multirow{3}{*}{ThisWork/Assef+18} &
  \multicolumn{3}{c|}{R75} &
  \multicolumn{3}{c|}{R90} \\ \cmidrule(lr){2-4} \cmidrule(lr){5-7} 
 &
  \multicolumn{1}{c|}{SPIRE} &
  \multicolumn{1}{c|}{Red} &
  \multicolumn{1}{c|}{Test} &
  \multicolumn{1}{c|}{SPIRE} &
  \multicolumn{1}{c|}{Red} &
  Test \\ \cmidrule(lr){2-4} \cmidrule(lr){5-7} 
 &
  \multicolumn{1}{c|}{0.38} &
  \multicolumn{1}{c|}{0.60} &
  \multicolumn{1}{c|}{0.95} &
  \multicolumn{1}{c|}{0.63} &
  \multicolumn{1}{c|}{0.78} &
  1.0 \\ \hline
\multirow{2}{*}{Assef+18/True} &
  \multicolumn{3}{c|}{R75} &
  \multicolumn{3}{c|}{R90} \\ \cmidrule(lr){2-4} \cmidrule(lr){5-7} 
 &
  \multicolumn{3}{c|}{1.11} &
  \multicolumn{3}{c|}{1.06} \\ \hline
ThisWork/True &
  \multicolumn{6}{c|}{1.06} \\ \hline
\end{tabular}
\end{table}

\subsection{AGN fraction and their implications in protoclusters}

When considering the `in' and `out' sources as PC members and field sources, respectively, the AGN fraction that we find in PCs is not significantly higher than the fraction measured in the field as found by other studies.
An important thing to keep in mind is that only a few PCs in our sample are confirmed (see Section \ref{sec:planck27}). Therefore, we may have sources that are line-of-sight alignments as suggested by \cite{Negrello2017}, instead of members of the overdensities, contaminating our sample.
We tested if measuring the AGN fraction in a subsample with a higher overdensity significance (i.e. a subsample of red sources), resulted in a higher AGN fraction. We found a higher AGN fraction than the field by at least factor 3 with 1$\sigma$ uncertainty. This could suggest that by selecting this redder sample we were in fact cleaning our sample and removing `line-of-sight alignments', and we would be consistent with higher AGN fraction in PCs.

Another possible explanation for finding an AGN fraction not as highly significant in PCs is that, as described in Section \ref{sec:planck27}, we are using a sample of PC candidates that are the most star-forming and dustiest members, instead of the full PC population, and the AGN population might not overlap with these.

Finally, the most likely explanation is that many PC members are too faint to be detected by WISE. Further, several PC AGN members may be detected only by W1 and W2 bands, and not in W3. In that case, this would mean that we are looking into the brightest AGN in the structures, which are rare. 
In order to test this statement, we looked into how many members of the protocluster PHz G237.01+42.50 (G237) at $z=2.16$ are detected by WISE. This PC has 31 spectroscopically confirmed members \citep{Polletta2021}.
Using a crossmatch radius of 6.5$\arcsec$ (W3 band resolution), out of the 31 sources, we found 5 WISE counterparts. For these counterparts, none of them are detectable in the W3 band, i.e. they have, on average, an SNR$<1$ in the W3 band. In other words, a $\sim$ 16 \% of the members were detected in W1 and W2 bands. Similarly, we consider the protocluster MAGAZ3NE J095924+022537 at z=3.37 \citep{McConachie2022}. Out of 22 spectroscopically confirmed members, we found 7 sources within 10$\arcsec$; none were detected in the W3 band. Thus, a $\sim$ 31 \% of the members were detected only in W1 and W2 bands.

Following this analysis, a diagnostic based only on the W1 and W2 may be considered for future work. In this case, we find that using W3 became a disadvantage in our method, and maybe other colours should be tested to find a better separation between AGN from star-forming galaxies, without biassing the sample to the most star-forming sources. Alternatively, a stacking analysis on the W3 signal could be done to reveal sources that are too faint to be detected individually.  
Also, our analysis could point to the fact that the small difference we found in the AGN fractions for field and PCs may be significant even if statistically is not. Thus, even if we did not find a highly significant difference, we think our results are still hinting at a higher AGN activity in PCs.

One of the main limitations of this study is that we are using photometrically selected PC candidates, instead of spectroscopically confirmed structures, due to the paucity of confirmed PCs available. Having a large data set of spectroscopically confirmed overdensities at high redshift would make it possible to better understand the relationship between AGN fractions $-$ and, therefore, the growth history of SMBHs in galaxies $-$ and the evolutionary state of early dense environments.

Nevertheless, WISE-selected AGN appear to be good indicators of overdensities \citep{Jones17}, as well as other AGN selections in general (e.g. \citealt{Noirot2016,Noirot2018}). Plus, follow-up observations from Spitzer/IRAC for some of these PC candidates \citep{martinache18}, continue to support the idea that these sources, or at least a good fraction of them, are true members of PC overdensities.

\section{Conclusions}
\label{sec:conclusions}

We estimated the AGN fraction in $\sim$228 protocluster candidates selected by Planck XXVII and followed up by Herschel \cite{PlanckCollaboration2015}, a representative sample of high redshift PC candidate members. This sample provides the photometry for 7099 sources and allows us to compare the measured AGN fraction of galaxies inside the overdensities and compare them with field galaxies.

We used the WISE counterparts of these sources since WISE provides higher-resolution photometry and the possibility of probing the stellar emission. This resulted in a catalogue of 646 counterparts. 

In order to select the AGN in our sample, we constructed a classifier based on a mid-IR AllWISE colour-colour selection criterion. This is achieved by combining W1-W2 > 0.94 and W2-W3 < 4.04 colour cuts, which corresponds to the mean minus 3$\sigma$ of the W1-W2 and mean plus 3$\sigma$ of the W2-W3 AGN distributions of a control sample made up of AGN and non-AGN catalogues, and a \textit{k-means} clustering algorithm that separates the control sample following the W1-W2 = 1.53(W2-W3) - 4.80 relation. Our control sample includes known AGN and non-AGN galaxies that were used to train our classifier.

Out of the 150 PC members, we found 17 AGN and 133 non-AGN, which corresponds to an AGN fraction of $f_{AGN_{in}}$ = 0.113 $\pm$ 0.03 or 11\% $\pm$ 3\%. For the 496 field galaxies, we found 47 AGN and 449 non-AGN, equivalent to an AGN fraction of $f_{AGN_{out}}$ = 0.095 $\pm$ 0.013 or 10\% $\pm$ 1\%.

For further study of the AGN fraction in PCs, we also measured the AGN fraction in a `redder' ($S_{350}/S_{250} > 0.7$ and $S_{500}/S_{350} > 0.6$) subsample of our SPIRE sources, which has a higher overdensity significance. In this case we consider the red sources as PC members and the non-red sources as field galaxies. We found an AGN fraction of $f_{AGN_{red}} = 0.186 \pm 0.044$ or $19\%$ $\pm$ 4\% and a $f_{AGN_{non-red}} = 0.037 \pm 0.010$ or $4\%$ $\pm$ 1\%. Moreover, to assess our AGN fraction for the field sample, we also measured the AGN fraction in the Lockman-SWIRE field from HerMES. We found an AGN fraction of $f_{AGN_{HerMES}} = 0.075 \pm 0.008$ or $8\%$ $\pm$ 1\%.

In terms of AGN activity in PCs, we found that our AGN fraction is not significantly higher in PCs when compared to the field, when considering the `in' and `out' sources as PC and field galaxies, respectively. For the field, we found that both our sample (`out') and the one from HerMES have a similar AGN fraction, thus suggesting that we have a representative field sample. However, we think that our results hint towards a higher SMBH activity in overdensities, specially since we found a higher difference in the AGN fraction for the red and non-red samples, which are proportional to the overdensity significance of the sample.

Our main conclusion is that it is complicated to assess the AGN and SMBH activity in overdensities, particularly at these high redshifts. We believe that it is necessary for a combined and complete multi-wavelength study to better understand the role of the environment in the evolution of galaxies and their SMBHs.
We expect that new observations from the James Webb Space Telescope will improve this kind of study by delivering deeper and higher resolution data for galaxies and large-scale structures in the redshift interval considered in this work.

\section{Acknowledgements}
We thank the anonymous referee for their helpful comments.

C.G. acknowledges funding by Universidad de Concepci\'on for the tuition scholarship. R.D. gratefully acknowledges support from the Chilean Centro de Excelencia en Astrof\'isica y Tecnolog\'ias Afines (CATA) Basal grant FB210003.

This publication makes use of data products from the Wide-field Infrared Survey Explorer, which is a joint project of the University of California, Los Angeles, and the Jet Propulsion Laboratory/California Institute of Technology, and NEOWISE, which is a project of the Jet Propulsion Laboratory/California Institute of Technology. WISE and NEOWISE are funded by the National Aeronautics and Space Administration.

\section{Data Availability}
The data underlying this article will be shared on reasonable request to the corresponding author.

\bibliographystyle{mnras}
\bibliography{refer}

%%%%%%%%%%%%%%%%% APPENDICES %%%%%%%%%%%%%%%%%%%%%

\appendix

\section{Monte Carlo results}\label{sect:monte-carlo}

\begin{figure*}
\begin{subfigure}{.5\textwidth}
    \centering
    \includegraphics[width = \textwidth]{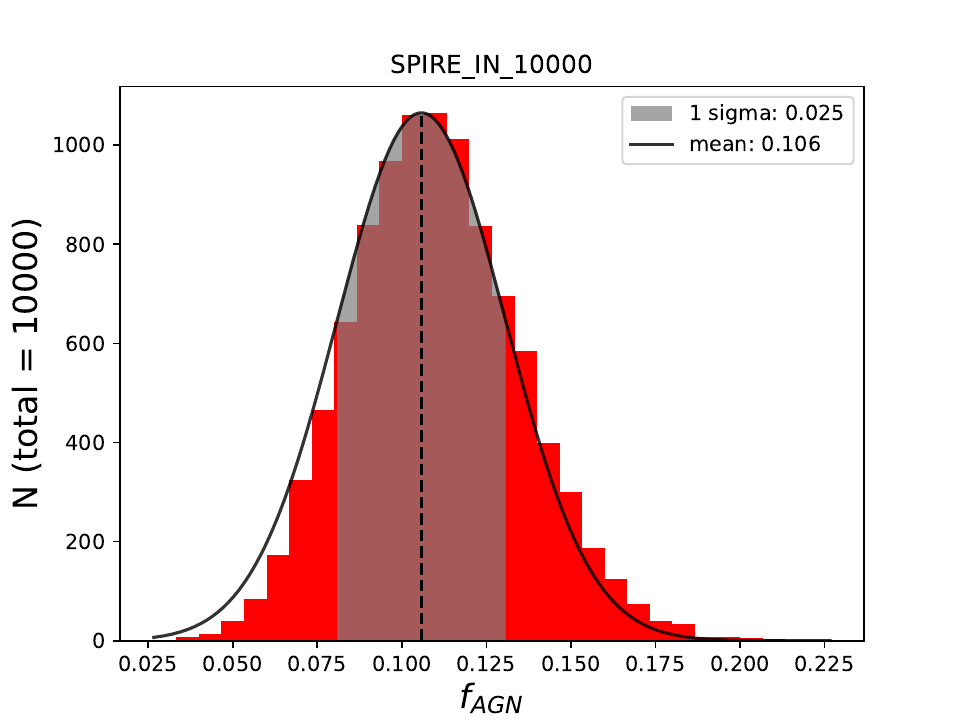}
\end{subfigure}~
\begin{subfigure}{.5\textwidth}
    \centering
    \includegraphics[width = \textwidth]{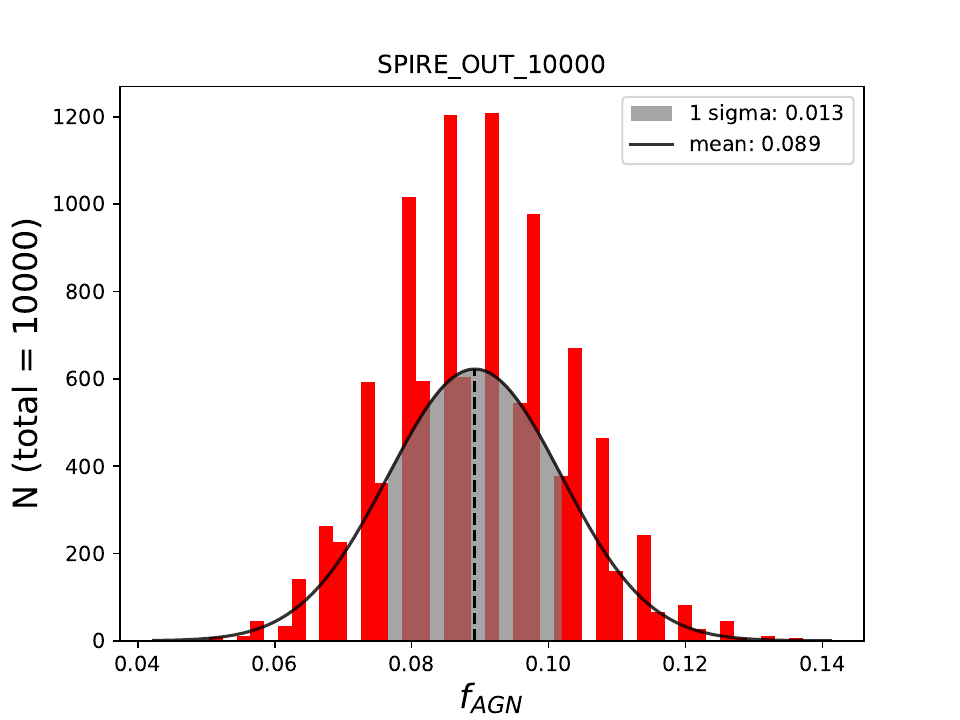}
\end{subfigure}
\begin{subfigure}{.5\textwidth}
    \centering
    \includegraphics[width = \textwidth]{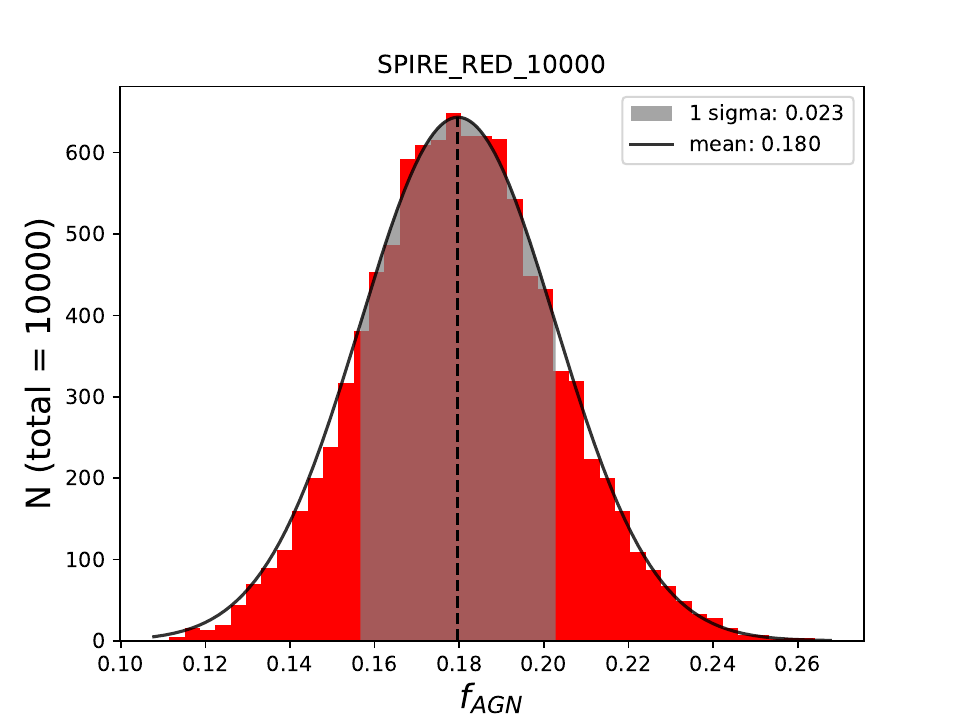}
\end{subfigure}~
\begin{subfigure}{.5\textwidth}
    \centering
    \includegraphics[width = \textwidth]{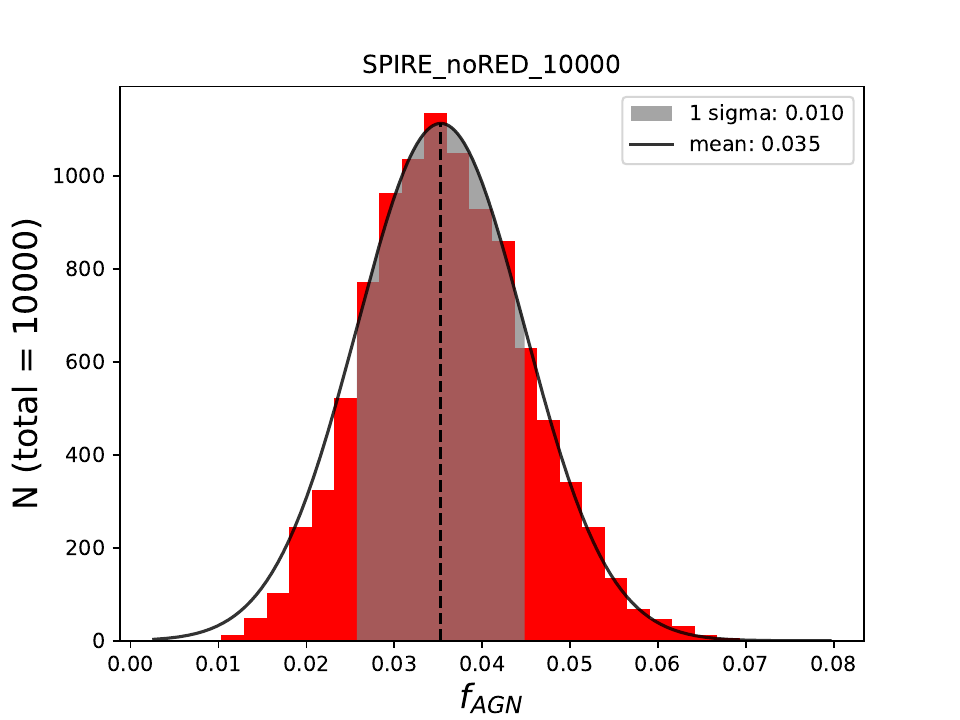}
\end{subfigure}
\begin{subfigure}{.5\textwidth}
    \centering
    \includegraphics[width = \textwidth]{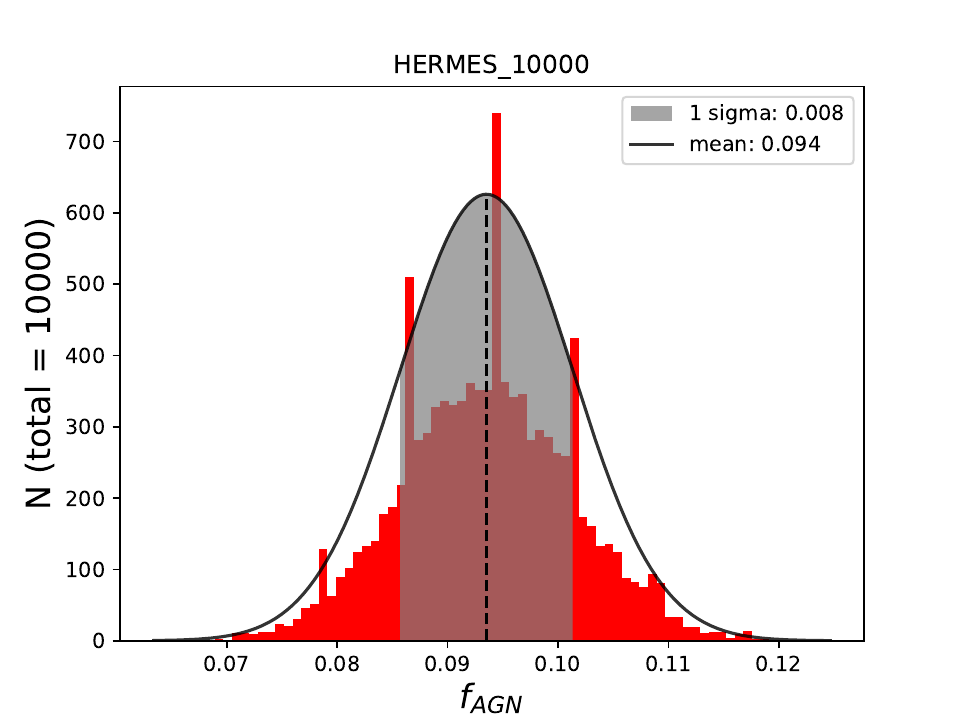}
\end{subfigure}
\caption{Resulting normal distributions (red) for the 10,000-step Monte Carlo simulation of the AGN fractions for sources inside (top left) and outside (top right) PCs. The normal distributions for the Monte Carlo simulations for our red (middle left) and non-red (middle right) subsamples and for the HerMES field sample (bottom centre) are also shown. 
The solid black line shows a fitted Gaussian model, with its corresponding mean (dashed black line) and standard deviation (shaded grey region). From these normal distributions, we use the 1$\sigma$ standard deviation as our AGN fractions uncertainties. This is, the AGN fraction for `in' sources has an error of 0.025, the AGN fraction for `out' sources has an error of 0.013, the AGN fraction of the red sample has an error of 0.023, the non-red sample has an error of 0.01 and the AGN fraction of the HerMES sample has an error of 0.008.}
\label{fig:arrs}
\end{figure*}

This appendix presents the normal distributions resulting from our Monte Carlo simulations. Figure \ref{fig:arrs} shows the distribution of the AGN fractions $f_{AGN}$ measured in each of the 10,000 iterations for the  `in' (top left), `out' (top right), red (middle left), non-red (middle right) and HerMES (bottom centre) sources. For each distribution we fitted a Gaussian model and found a mean of 0.106, 0.089, 0.180, 0.035 and 0.094  and a standard deviation of 0.025, 0.013, 0.023, 0.010 and 0.008 for the `in', `out', red, non-red and HerMES sources, respectively. We use the 1$\sigma$ standard deviation as our AGN fraction uncertainties.

\label{lastpage}
\end{document}